\documentclass[aps,prl,twocolumn,amsmath,amssymb,showpacs,superscriptaddress,floatfix,nofootinbib,]{revtex4}

\usepackage{graphicx}% Include figure files
\usepackage{dcolumn}% Align table columns on decimal point
\usepackage{bm}% bold math
\usepackage{mathrsfs}
\usepackage{appendix}
\usepackage{bbm}
\usepackage{color}

\def \vH {{\bf B}}

\def \vS {{\bf S}}

\def \vm {{\bf m}}

\def \mb {\mu_{\rm B}}
\def \hsf {B_{\rm SF}}

\def \va {{\bf a}}
\def \vb {{\bf b}}
\def \vc {{\bf c}}

\def \mnc {{\rm MnCl$_3$(bpy)} }

\begin{document}

\title{{Long-range magnetic order and interchain interactions in the $S=2$ chain system 
MnCl$_3$(bpy)}\footnote{Copyright notice: This manuscript has been authored by UT-Battelle, LLC under Contract 
No.~DE-AC05-00OR22725 with the U.S. Department of Energy. The United States Government retains and the publisher, 
by accepting the article for publication, acknowledges that the United States Government retains a non-exclusive, 
paid-up, irrevocable, world-wide license to publish or reproduce the published form of this manuscript, or allow 
others to do so, for United States Government purposes. The Department of Energy will provide public access to 
these results of federally sponsored research in accordance with the DOE Public Access Plan 
(http://energy.gov/downloads/doe-public-access-plan).}}

\author{Randy S.~Fishman}
\affiliation{Materials Science and Technology Division, Oak Ridge National Laboratory, Oak Ridge, Tennessee 37831, USA}
\author{Shin-ichi Shinozaki}
\affiliation{Center for Advanced High Magnetic Field Science, Graduate School of Science, Osaka University, 
Toyonaka, Osaka 560-0043, Japan}
\author{Akira Okutani}
\affiliation{Center for Advanced High Magnetic Field Science, Graduate School of Science, Osaka University, 
Toyonaka, Osaka 560-0043, Japan}
\author{Daichi~Yoshizawa}
\affiliation{Center for Advanced High Magnetic Field Science, Graduate School of Science, Osaka University, 
Toyonaka, Osaka 560-0043, Japan}
\author{Takanori Kida}
\affiliation{Center for Advanced High Magnetic Field Science, Graduate School of Science, Osaka University, 
Toyonaka, Osaka 560-0043, Japan}
\author{Masayuki Hagiwara}
\affiliation{Center for Advanced High Magnetic Field Science, Graduate School of Science, Osaka University, 
Toyonaka, Osaka 560-0043, Japan}
\author{Mark W.~Meisel}
\affiliation{Department of Physics and the National High Magnetic Field Laboratory, University of Florida, 
Gainesville, Florida 32611-8440, USA}
\affiliation{Joint Institute for Neutron Sciences, Oak Ridge National Laboratory, Oak Ridge, TN 37831-6453, USA}

\date{\today}

\begin{abstract}
A compound with very weakly interacting chains, MnCl$_3$(bpy), has attracted a great deal of attention as a possible $S=2$ Haldane chain.  
However, long-range magnetic order of the chains prevents the Haldane gap from developing below 11.5~K.   
Based on a four-sublattice model, a description of the antiferromagnetic resonance (AFMR) spectrum up to frequencies of 1.5~THz 
and magnetic fields up to 50~T indicates that the interchain coupling is indeed quite small but that the
Dzaloshinskii-Moriya interaction produced by broken inversion symmetry is substantial (0.12~meV).  
In addition, the antiferromagnetic, nearest-neighbor interaction within each chain (3.3~meV) is
significantly stronger than previously reported.  The excitation spectrum of this 
$S=2$ compound is well-described by a $1/S$ expansion about the classical limit.
\end{abstract}

\pacs{76.50.+g, 75.10.Jm, 75.50.Ee}

\maketitle

\section{Introduction}

Magnetic chains composed of spin $S = 2$ ions have received considerable theoretical and numerical attention 
\cite{Tonegawa2011,Tu2011,Pollmann2012,Tzeng2012,Hasebe2013,Kjall2013,Chen2015,Kshetrimayum2015}
due to their unique predicted behavior.   
Even- and odd-integer spin chains are distinct, with the latter in a symmetry-protected 
topological phase \cite{Pollmann2012,Pollmann2010}.  
Whereas the $S=1$ Haldane phase \cite{Haldane1,Haldane2,Schulz1986,Affleck1989} has been 
observed experimentally \cite{Renard,Hagiwara1990,Glarum1991,Avenel1,Avenel2,Yamashita,Affleck-staggered},  
formation of the $S=2$ Haldane state has been prevented by long-range magnetic ordering due to interactions between the chains.  
Although some aspects of $S=2$ chains have been observed in optical gasses \cite{Chen2015,Lewenstein2007}, 
the question remains whether a real chain can realize the $S=2$ Haldane phase 
\cite{Matsuhita2003,Leone2004,CizmarJMMM,Stock2009,Stone2013,Cizmar2016}.

\begin{figure}[b!]
\begin{center}
\includegraphics[width=3.375in]{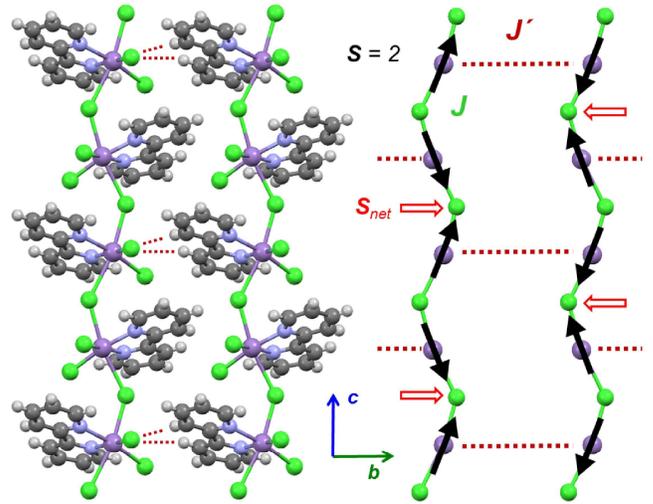}
\end{center}
\caption{(Color online) The left side shows the crystal structure of MnCl$_3$(bpy) \cite{Perlepes} for two nearest-neighbor 
chains in the $b-c$ plane.  The staggered chains of Mn(III) $S=2$ ions are connected by Cl atoms, and the locations of the (bpy) 
cause an alternating Cl$\cdot\cdot\cdot$H coupling indicated by the dotted lines.  The right side shows only the Mn$-$Cl chains  
and the interactions $J$ and $J^{\prime}$. The Mn magnetic moments are indicated by the dark arrows, while open arrows sketch 
the net moments arising from the canted spins.}
\label{fig1}
\end{figure}

Portrayed in Fig.~\ref{fig1}, our protagonist is the $S=2$ antiferromagnetic chain of ($2,2^{\prime}$-bipyridine)trichloroman- ganese(III), 
MnCl$_3$(bpy), where (bpy) = ($2,2^{\prime}$-bipyridine) = C$_{10}$H$_8$N$_2$, \cite{Goodwin,Perlepes}.
Due to the (bpy) molecules
separating the chains, this material was believed to be an excellent candidate for observing the $S=2$ Haldane phase \cite{Granroth1996}.
However, weak signatures from randomly-arranged microcrystals hinted that long-range order might appear at low temperatures 
\cite{Hagiwara2012,Hagiwara2013}.  Recently, unambiguous long-range antiferromagnetic ordering 
was identified at $T{_\mathrm{N}} = 11.5$~K \cite{Hagiwara2015,Shinozaki2016} in oriented single-crystals.  

Although magnetic ordering appears in
single crystals, the recently published antiferromagnetic resonance (AFMR) spectra of 
MnCl$_3$(bpy) \cite{Shinozaki2016} was not accurately described by a quasi-classical, two-sublattice calculation for isolated (non-interacting) 
chains \cite{Kanamori,Date,Hagiwara-review}.  Nevertheless, those results suggest that the classical Heisenberg model  
is an appropriate starting point for the Hamiltonian, which needs to also include other important interactions  
such as the Dzaloshinskii-Moriya (DM) interaction between neighboring spins and the exchange coupling between adjacent 
chains \cite{Huang2004,Herak2013}.  

The forthcoming analysis provides an excellent description of the magnetic field dependences of the AFMR mode frequencies in the presence 
of a sizable DM term.  Strikingly, only an extremely weak interchain coupling is required to drive long-range 
antiferromagnetic order.  Consequently, the $S=2$ Haldane phase is unlikely to be detected in molecule-based magnets.

Our more sophisticated analysis of the AFMR spectrum includes both intrachain and interchain couplings $J$ and $J^{\prime}$, 
respectively, as well as the DM interaction $D$ generated by broken inversion symmetry.  
From Fig.~\ref{fig1}, the DM interaction vector lies along the $\pm \, \va ^* $ directions, alternating in sign along each chain.  Our
description also includes the easy-axis anisotropy $K$, which favors spin alignment along the chain axis $c$ ($K >0$) or in the $a^*-b$ plane ($K < 0$),
and the easy-plane anisotropy $E$, which favors spin alignment along $\vb $ ($E > 0$) or along $\va^*$ ($E < 0$).  
As found earlier, 
the $\underline{g}$ tensor will be taken to be slightly anisotropic with eigenvalues 
$g_{a^*a^*} = 2.09$, $g_{bb} = 1.92$ and $g_{cc}=2.07$ \cite{Hagiwara2013,Shinozaki2016}.
Note that we have modified the previous notation \cite{Shinozaki2016}, where $D$ was used to represent 
the single-ion anisotropy along $c$, now defined as $K$.

\section{Experimental details}

The high-field magnetization of some single crystal samples of MnCl$_3$(bpy) along the $c$ axis was measured again in pulsed magnetic fields up to 47~T using a standard induction 
method with a pick-up coil arrangement.  The signal response was calibrated by comparison with the data obtained with the SQUID magnetometer up to 7~T. 
High-field, multi-frequency electron-spin resonance data were taken from Ref.[\onlinecite{Shinozaki2016}], where details of the sample preparation are given. 
Due to sample deterioration, the extrinsic magnetization was subtracted from the raw data to get the intrinsic magnetization curve by assuming a $S=5/2$ Brillouin function
as in Ref.[\onlinecite{Hagiwara2012}].
The subtracted magnetization at 4.2 K  and below 7 T then coincided with the magnetization measured previously with the SQUID magnetometer.
The maximum error bar in the magnetization at 40 T is $\pm$ 10 $\%$. 

\section{Model}

With magnetic field $\vH $ along $\vm $, the Hamiltonian of \mnc can be written as
\begin{eqnarray}
{\cal H}&=&-J\sum_{i, k} \vS_i^{(k)} \cdot \vS_{i+1}^{(k)} -J^{\prime} \sum_{i, k} \vS_i^{(k)} \cdot \vS_i^{(k+1)} \nonumber \\
&-& K \sum_{i,k} S_{iz}^{(k) 2} + E \sum_{i, k} \Bigl( S_{ix}^{(k) 2} -S_{iy}^{(k) 2} \Bigr) \nonumber \\
&-& D\sum_{i, k} (-1)^i\, \va ^* \cdot \Bigl(\vS_i ^{(k)}\times \vS_{i+1}^{(k)} \Bigr) \nonumber \\
&-& \mb B \sum_{i ,k}  \vm \cdot \underline{g} \cdot \vS_i^{(k)}\,\,\,,
\label{eq1}
\end{eqnarray}
where the chain index is given by $k$ and the site index on each chain is given by $i$.  The direction of the DM vector $\bf D$ along
$\va ^*$ was chosen to conform with the symmetry rules provided by Moriya \cite{Moriya1960} for materials with broken inversion symmetry.  
The factor $(-1)^i$ in front of the DM interaction reflects the alternation in the position of the (bpy) radical along the chain.  
We take $J < 0$ and $J' < 0$ for antiferromagnetic couplings. 

The magnetic ground state of this Hamiltonian is obtained by minimizing the energy $\langle \cal H \rangle $ 
for the 8~angles of the four classical spins that form the magnetic unit cell, and the excitation spectrum 
is obtained by performing a $1/S$ expansion about the classical limit.  Assuming a linear response for weak 
perturbation from equilibrium, solving the equations-of-motion 
requires the numerical diagonalization of a $8 \times 8$ matrix.

An earlier study of the AFMR excitation spectrum neglected both $J^{\prime}$ and $D$ \cite{Shinozaki2016}, 
while the value for the nearest-neighbor coupling $J$ $(-2.69$~meV = $-31.2$~K) was estimated from the peak 
in temperature-dependence of the low-field magnetic susceptibility assuming $K = 0$ \cite{Hagiwara2013}.  Using their
values for the parameters (Table~\ref{table1}), the caclulated mode frequencies in Fig.~\ref{fig2}(a) reproduce the ones reported by 
Shinozaki \emph{et al.}~\cite{Shinozaki2016}.  In general, the experimental spectra are satisfactorily represented 
by those calculations, but the 10\% overestimation of the spin-flop field $B_{\mathrm{SF}}$ and 
the error in the mode frequencies for $\vm = \vc$ and $B > B_{\mathrm{SF}}$ are troubling issues.

\begin{table}[t!]
\caption{Exchange and anisotropy parameters in meV (uncertainties discussed in text).}
\begin{ruledtabular}
\begin{tabular}{ccccccc}
& $J$ & $J^{\prime }$ & $K$  & $E$ & $D$ & $\chi^2$
  \\
 \hline
Ref.[\onlinecite{Shinozaki2016}] & $-2.69$ & 0 & 0.129 & 0.015 & 0 & 0.211\\ 
This work & $-3.3$  & 0  &  0.102  & 0.018 & 0.12 & 0.035  \\
uncertainties & $\pm 0.4$ & $\pm 0.001$ & $\pm 0.014$ & $\pm 0.003$ & $\pm 0.04$ &  \\ 
\end{tabular}
\end{ruledtabular}
\label{table1}
\end{table}

\section{Numerical Fits}

Due to the uncertainty in $J$, the other parameters in the Hamiltonian of Eq.(\ref{eq1}) are calculated by fitting the AFMR data with fixed $J$.  
For $J < -2$~meV, the best fits are always obtained as $J^{\prime} \rightarrow 0$.  Of course, a small negative 
(antiferromagnetic) $J^{\prime}$ is required to cancel the moments on adjacent chains.  
The result of this analysis over a range of $J$ values is shown in Fig.~\ref{fig3}(a), where 
the DM coupling constant $D$ becomes markedly smaller as $\vert J\vert$ decreases.  
For fixed $J$, the statistical uncertainties in $J'$, $K$, $D$, and $E$ are evaluated from the variation in $\chi^2$.
The anisotropies $K$ and $E$ are always positive, corresponding to one easy axis along $\vc $ and a
second easy axis along $\vb$.  Both anisotropies grow as $\vert J \vert $ decreases.

\begin{figure}[ht!]
\includegraphics[width=8cm]{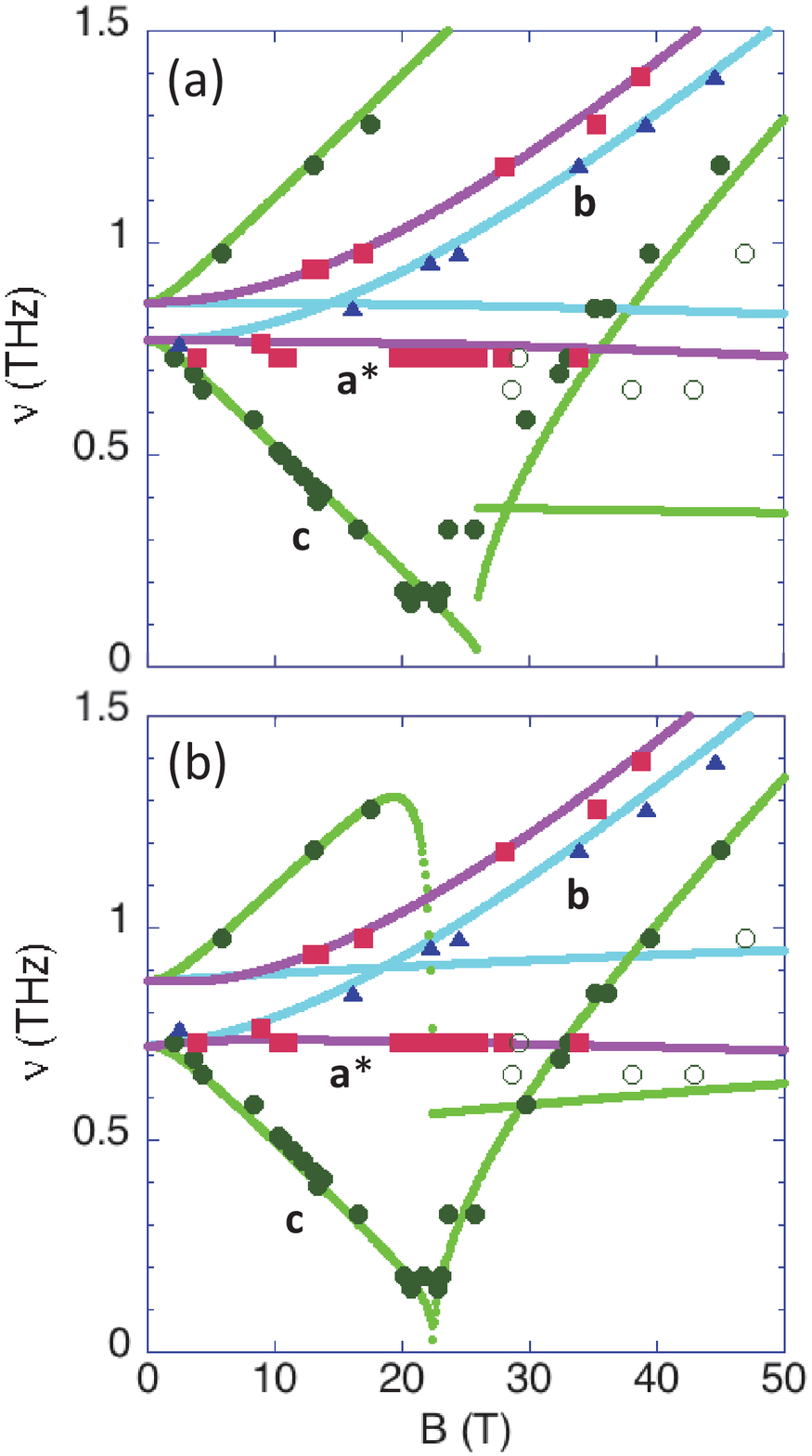}
\caption{(Color online) The magnetic field dependences of the AFMR frequencies of MnCl$_3$(bpy) for $T \approx 1.3$~K 
($\vm || \va^*$ and $\vm || \vb$) or 1.3, 1.5, and 1.7~K ($\vm || \vc$).
The data points are from the experimentally observed resonances \cite{Shinozaki2016} for magnetic field $B\vm $ applied 
parallel to $\va^*$ (red squares), $\vb $ (blue triangles), and $\vc $ (green circles).  
(a) The lines are the results of the calculations reported by Shinozaki \emph{et al.}~\cite{Shinozaki2016} and 
reproduced here with the values for the parameters listed in Table~\ref{table1}.  (b) The results of this work 
using Eq.(\ref{eq1}) and the analysis presented in Fig.~\ref{fig3} to determine the parameters given in 
Table~\ref{table1}.  Open circles are ``outlier" points for the field along the $c$-axis (see discussion in the text).} 
\label{fig2}
\end{figure}   

The $\chi^2$ value of the fits decreases from 0.0383 at $J=-5$~meV to a minimum of 0.0306 at $J=-2.3$~meV, as shown in Fig.~\ref{fig3}(a).
Because all $\chi^2$ values in this range of $J$ are acceptable, we use the magnetization as an additional constraint on $J$.
The $a^*$-axis, $b$-axis, and $c$-axis magnetizations at 40~T are calculated 
as a function of $J$ and plotted in Fig.~\ref{fig3}(b).  Since the magnetization is a function of $\mb B /\vert J\vert $, 
a smaller value of $\vert J\vert $ enhances both the effective field and the magnetization.  Notice that the 
predicted values of $M_{a^*}$ and $M_b$ are quite close and cross at $J=-2.8$ meV.
The experimental value for the magnetization $M_b^{\rm exp} \approx 0.68\, \mb$ with field along 
$\vb $ is also indicated in this figure \cite{Shinozaki2016}.  

Figure~\ref{Mag_c} shows earlier magnetization curves~\cite{Shinozaki2016} at 1.7~K along the $a^*$ and $b$ directions.  
The curve at 1.4~K along the $c$ axis was remeasured 
to check the large deviation of the earlier measurements from the calculated magnetization. 
As before~\cite{Shinozaki2016}, the magnetization curve for $\vm || \vc $ indicates a spin-flop transition at 22~T.
Above this spin-flop field, the slope of the magnetization curve is larger than previously reported because the sample allignment 
along the $c$ axis has now been corrected. 

Based on $M_b^{\rm exp}$, the best value for the nearest-neighbor interaction is $J \approx -3.60$~meV. 
However, our new results indicate that $M_c^{\rm exp} \approx 0.9 \, \mb $, suggesting that $J \approx -2.95$~meV.
It is important to recognize that these values reflect anisotropy contributions that were
neglected in the earlier estimate $J \approx -2.69$~meV \cite{Shinozaki2016}.

So comparison with the experimental magnetization suggests that $J = -3.3\pm 0.4$~meV.
The corresponding anisotropy and DM parameters from Fig.~\ref{fig3} are $K = 0.102 \pm 0.014$~meV, $E= 0.018 \pm 0.003$~meV, 
and $D=0.12 \pm 0.04$~meV.   Within an uncertainty of $\pm 1.3 \times 10^{-3}$~meV, $J'$ is zero.
All parameters and their uncertainties are given in Table~\ref{table1}.  
Compared with earlier fits \cite{Shinozaki2016}, 
$K $ is smaller but $E$ is slightly larger. 

\begin{figure}
\includegraphics[width=8cm]{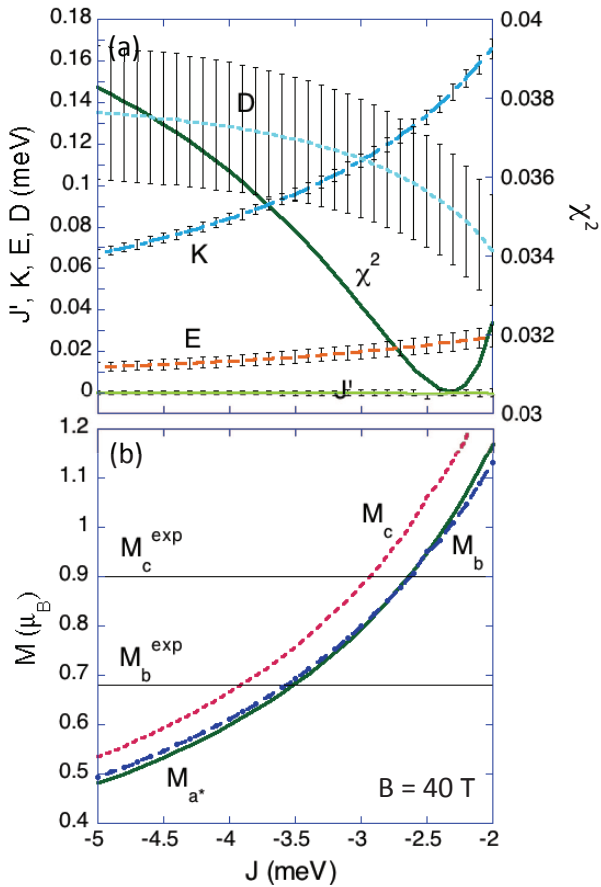}
\caption{(Color online) (a) The anisotropy, DM interactions, $\chi^2$, and (b) the magnetizations at 40 T versus $J$.  
Horizontal lines in (b) shows the experimental magnetizations for a 40 T field (see text) along $\vb $ or $\vc $.  
}
\label{fig3}
\end{figure}

The value $D=0.12$~meV for the DM interaction corresponds to a tilt 
of each spin at zero field by about 1$^{\circ}$ towards the $b$-axis.  This canting is associated with a net moment
${\bf M}_{\rm net }\approx \pm 0.07 \, \mb \, \vb $, alternating in sign on neighboring chains.  
The new fits provide a $\chi^2$ value about 6~times smaller than the fits in 
Ref.[\onlinecite{Shinozaki2016}].  The five points indicated by open circles in 
Fig.~\ref{fig2}, all obtained with field along $\vc $, are not included in this analysis.  These points seem to be ``outliers'' with respect to the main $c$-axis mode for 
$B > B_{\mathrm{SF}}$ and may be associated with other flat branches due to a small misalignment of the crystal.  Including these ``outliers"
would increase $\chi^2$ but would not change the fitting parameters in Table~\ref{table1}. 

\begin{figure}
\includegraphics[width=8cm]{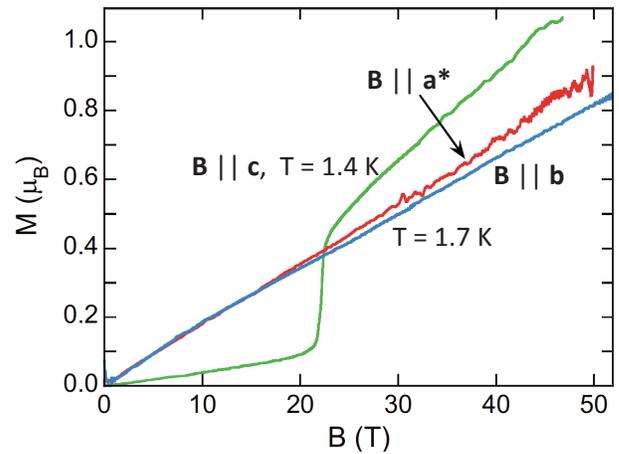}
\caption{(Color online) The magnetization curves for field along the $a^*$, $b$, or $c$ crystalline axes.  
The magnetization along the $c$ axis was remeasured and those along the other directions are taken from Ref.[\onlinecite{Shinozaki2016}].
}
\label{Mag_c}
\end{figure}

The resulting fits to the AFMR spectrum are plotted in Fig.~\ref{fig2}(b), where excellent agreement now exists 
between the calculated value for $B_{\mathrm{SF}} = 22.4$~T and the experimentally determined one.  
In addition, the predicted mode frequencies are in much better agreement with the measured 
mode frequencies when $\vm = \vc $.  The lower predicted mode frequency reaches a minimum of about $3\times 10^{-3}$ THz 
at $\hsf $, and it is noteworthy that both branches of the excitation spectrum soften as $B$ approaches $\hsf$.  
Aside from $\vm = \va^* $, the other predicted ``flat" modes are too weak to be observed, but 
they are included in Fig.~\ref{fig2} for completeness.

\section{Conclusion}

Surprisingly, the expansion about the classical limit or linear spin-wave theory works very well for this putative quantum-spin system.
Since $J^{\prime} /J  \lesssim 4 \times10^{-4}$, the coupling between chains is very weak in MnCl$_3$(bpy).
Nevertheless, the ordering temperature of MnCl$_3$(bpy) is about 11.5 K [\onlinecite{Hagiwara2015,Shinozaki2016}].
For a quasi-two-dimensional system with small exchange $J'$ between planes, the critical temperature scales like
$\vert J\vert {\rm{log} }(J'/J)$ \cite{Yasuda2005}.  For a two-dimensional antiferromagnet with easy-axis anisotropy $K$, the critical temperature scales 
like $\vert J\vert {\rm{log}} (K/\vert J\vert )$ \cite{Yosida1996}. 
Since no long range order is possible in one dimension, even with anisotropy, it is unclear how the critical temperature scales with 
$J^{\prime }/J $.   If $T_N$ scales like $\vert J\vert {\rm{log} }( J'/J) $, then
even a very small value of $J^{\prime}$ can stabilize long-range magnetic order with a N\'eel temperature of 10 K.
If instead, $T_N$ scales like $\sqrt{J^{\prime }J}$, then $J'=4\times 10^{-4} J$ would correspond to a mean-field N\'eel temperature of about 6 K
in the absence of anisotropy.  Either scaling may explain the magnetic ordering in MnCl$_3$(bpy).

To summarize, we have used linear spin wave theory to obtain an excellent description of the AFMR spectrum in MnCl$_3$(bpy).  
Since an expansion about the classical limit works very well for MnCl$_3$(bpy), researchers searching for 
an $S=2$ Haldane chain should explore other options.

\begin{acknowledgements}

We thank Sasha Chernyshev for useful conversations.
Research sponsored by the Department of Energy, Office of Science, Basic Energy Sciences, Materials Sciences and Engineering Division (RSF),
by Grants-in-Aid for Scientific Research (Grants No.~24240590, No.~25246006, and No.~25220803) from the MEXT, Japan (MH), and by the
National Science Foundation through Grant No. DMR-1202033 (MWM).  A part of this work was conducted under the foreign visiting professor 
program of the Center for Advanced High Magnetic Field Science.

\end{acknowledgements}

\bibliography{MnCl3bpy}

\end{document}